\documentclass[aps,epsf,amsfonts,floats,twocolumn,amssymb,amsmath,groupedaddress,showpacs,floatfix,nofootinbib]{revtex4}
\usepackage{graphicx}
\usepackage{epsf}

\usepackage[]{latexsym}
\usepackage{bm}
\usepackage{amsmath}

\newcommand{\be}{\begin{equation}}\newcommand{\ee}{\end{equation}}
\newcommand{\bea}{\begin{eqnarray}}\newcommand{\eea}{\end{eqnarray}}
\newcommand{\brr}{\begin{array}}\newcommand{\err}{\end{array}}
\newcommand{\bit}{\begin{itemize}}\newcommand{\eit}{\end{itemize}}
\newcommand{\ben}{\begin{enumerate}}\newcommand{\een}{\end{enumerate}}

\newcommand{\ba}{\begin{array}}
\newcommand{\ea}{\end{array}}

\def\1{{_{1}}}\def\2{{_{2}}}

\def\noHe0{:\;\!\!\;\!\!:H_e(0):\;\!\!\;\!\!:}
\def\noHm0{:\;\!\!\;\!\!:H_\mu(0):\;\!\!\;\!\!:}

\def\1{{_{1}}}\def\2{{_{2}}}

\begin{document}

\title{The effect of the behavior of an average consumer on the public debt dynamics}

\author{Roberto De Luca$^1$}

\author{Marco Di Mauro$^1$}

\author{Angelo Falzarano$^2$}

\author{Adele Naddeo$^3$}

\affiliation{ $^1$ Dipartimento di Fisica E.R.Caianiello, Universit\'a di Salerno, Fisciano (SA) - 84084, Italy}

\affiliation{ $^2$ Dipartimento di Scienze Economiche e Statistiche, Universit\'a di Napoli Federico II, Napoli - 80126, Italy}

\affiliation{ $^3$ INFN Sezione di Napoli, Napoli - 80126, Italy}


\begin{abstract}

An important issue within the present economic crisis is understanding the dynamics of the public debt of a given country, and how the behavior of average consumers and tax payers in that country affects it. Starting from a model of the average consumer behavior introduced earlier by the authors, we propose a simple model to quantitatively address this issue. The model is then studied and analytically solved under some reasonable simplifying assumptions. In this way we obtain a condition under which the public debt steadily decreases.

\end{abstract}

\maketitle

\section{Introduction}

Understanding consumption is a key issue in macroeconomics. The links between individual consumption decisions and outcomes for the whole economy have been deeply investigated by A. Deaton \cite{DeatonBook}, who was awarded the Nobel prize in 2015 \cite{NobelPrize}. In fact, since the beginning, research focused on the relation between income (and the interest rate) and consumption. Within this framework, Friedman's permanent income hypothesis \cite{friedman} and Modigliani's life-cycle theory \cite{modigliani1, modigliani2, modigliani3} contributed to suggest how income variables might enter the consumption function. In particular life-cycle theory provides a useful framework for thinking about saving and related mechanism, and paves the way for undestanding the link between saving and growth of a country. Furthermore, the theory predicts that saving rates are higher in more rapidly growing economies, even in the short run, therefore establishing a correlation between growth and savings rates. However, more recent studies pointed out that the range of applicability of the above mechanism is quite limited \cite{study1, study2, study3, study4}, so that it alone does not succeed in explaining the international correlation between saving and economic growth.

Macroeconomists and policy-makers have traditionally been concerned with the issue of the
sustainability of public debt in developing and emerging market countries. However, starting from the appearance of the global financial crisis, the attention has been shifted to developed economies, which suffer from rising
debt-Gross Domestic Product (GDP) ratios in the face of stagnant or contracting output. As a matter of fact, in most European countries debt is at an unprecedented level in the last fifty years. In some cases, the increases since 2007 have exceeded 20 percentage points of GDP while the level of the debt was already high \cite{consolidation1}.

The correct definition and measurement of public debt were given by \cite{barro, DeHaan}. The relation between public debt and economic growth is the object of much work today \cite{GreinerFincke, Herndon, Panizza, Smyth, Nastanski1, Reichlin}. If debt growth rates are lower than those of GDP, the debt is not bad. The economic growth restores the income part of the budget, which is used to pay interest on debt. Vice versa, with low economic growth rates, public debt becomes a serious macroeconomic problem for the country. In such a case governments have to pay a large amount of public revenue to the creditors, which results in fewer resources for education and public investments. An unsustainable level of debt is obtained in correspondence to a high debt to GDP ratio \cite{Nastanski1}. In particular, interest rates rise on the basis of a greater probability of a default on debt obligations, with a negative influence on private investment and consumption. The debt-relief Laffer curve \cite{Miles} provides information about the negative impact of high debt on output growth; in particular a point can be found, where outstanding debt is so large that output growth gets reduced and the probability of debt repayment lowers. In this scenario, inflation could be a viable solution to the government debt problem \cite{inflation1, inflation2}, while the effectiveness of fiscal consolidation and of austerity requirements, such as the reduction in governments expenditure, is widely debated (see e.g. \cite{consolidation1} and references therein).

Actually, new provisions in the Stability and Growth Pact (SGP) require European countries with a debt to GDP ratio higher than 60 per cent to act to decrease it in the next few years. It is thus important to assess under which conditions this is possible. This is the purpose of the present paper, in which a simple model is proposed to try to find such conditions.

Since the model uses the representative agent framework, it is subject to all its limitations, as it does not take into account the differences between individuals. The use of the representative agent in economy has been criticized mainly in \cite{Kirman1992}. Progress in providing alternatives to the ad hoc assumptions of the representative agent approach has recently been made by using the tools of Statistical Mechanics (see e.g. \cite{DeMartinoMarsili1} and references therein).

Another related limiting aspect of the present analysis concerns the absence of private enterprises in this scenario. In fact, the presence of this type of agent can generate earnings that can substantially help in paying back public debt. However, if we assume that the economic behavior of private enterprises can be included in the representative agent picture, the present model could be still adopted in this more general setting.

In a previous paper \cite{noi} the authors proposed a simple mathematical model, based on a hydrodynamic analog, to quantitatively describe the time evolution of the amount of money an average consumer decides to spend, depending on his/her available budget. In the present paper we couple this model, or better, its difference equation version, to the standard equation for the time evolution of the public debt given in ref. \cite{barro}. The resulting dynamical model is then analytically solved under some reasonable simplifying assumptions. The result is a range of the parameters of the model for which the public debt steadily decreases. The model is presented in Section 2, and its dynamics is studied in Section 3. Finally, in Section 4, some conclusions are drawn, and some comments and perspectives of our work are outlined.

\section{The model}

The typical strategy adopted by a Government to decrease its public debt is to burden the average consumer with additional taxation, which may also be applied to the amount of money present, at a certain date, on the consumer's bank account.
Let us write the tax payed by a single average consumer in the k-th year by
\begin{equation}\label{tax}
\tau_k=\alpha p_a + \beta \delta_{km}b_k + \gamma c_k,
\end{equation}
where $p_a$ is the yearly income, which we assume fixed, $b_k$ is the average yearly bank deposit, and $c_k$ is the yearly expenditure of the consumer. $\alpha$, $\beta$ and $\gamma$ are the average taxation rates for these quantities, respectively. The Kronecker delta $\delta_{km}$ picks a fraction of the amount $b_k$ only in the m-th year.

The dynamics of the public debt $D_k$. calculated at the year $k$ and normalized to the number $N$ of consumers, is described by the difference equation \cite{barro}
\begin{equation}\label{debtdynamics}
g_k+rD_{k-1}=\tau_k + D_k -D_{k-1},
\end{equation}
where $D_{k}-D_{k-1}$ is the yearly pro capite deficit at the time $k$, $r$ is the fixed rate of return on public and private debts, $g_k$ is the pro capite amount of money the State utilizes for funding education, health, welfare, and all public activities. Plugging (\ref{tax}) into (\ref{debtdynamics}) we get
\begin{equation}\label{debtdynamics2}
D_k-(1+r)D_{k-1}=g_k-\alpha p_a -\beta b_k \delta_{km} -\gamma c_k
\end{equation}
To describe the behavior of the consumer we may adopt the leaking bucket model \cite{noi}:
\begin{equation}\label{leakingbucket}
b_k-b_{k-1}=y_k - c_k
\end{equation}
where $y_k$ is the expendable part of the salary, i.e. $y_k=p_a-\tau_k$. Substituting this definition into (\ref{leakingbucket}), we get
\begin{equation}
p_a-(\alpha p_a + \beta \delta_{km} b_k + \gamma c_k)-c_k=b_k - b_{k-1}
\end{equation}
Substituting $c_k=a b_k^2$ as in \cite{noi}\footnote{we assume that all consumers have no debts, i.e. a positive budget, cfr. \cite{noi}} we get
\begin{equation}\label{consumer}
(1+\gamma)a b_k^2 + (1+\beta\delta_{km})b_k-b_{k-1}=(1-\alpha)p_a.
\end{equation}
This equation should of course be coupled to the public debt evolution equation (\ref{debtdynamics2}). Let us now study the latter with the simplifying assumption $g_k=g_0=const.$ and the functional relation $c_k=a b_k^2$:
\begin{equation}
D_k-(1+r)D_{k-1}=(g_0-\alpha p_a) -\beta b_k\delta_{km}-\gamma a b_k^2
\end{equation}
or
\begin{equation}
D_k=(1+r)D_{k-1}+\Delta_k
\end{equation}
where we defined $\Delta_k=(g_0-\alpha p_a) -\beta b_k\delta_{km}-\gamma a b_k^2$. The solution of this difference equation is
\begin{equation}
D_k=(1+r)^k\left[D_0+\sum_{i=1}^k\frac{\Delta_i}{(1+r)^i}\right]
\end{equation}
The sum in the above equation is
\begin{eqnarray}\label{Coupon}
S_k&=&\sum_{i=1}^k\frac{\Delta_i}{(1+r)^i}\\&=&(g_0-\alpha p_a)\frac{(1+r)^k-1}{r(1+r)^k}-\gamma a \sum_{i=1}^{k}\frac{b_i^2}{(1+r)^i}-\frac{\beta b_m}{(1+r)^m}\nonumber
\end{eqnarray}
so that
\begin{eqnarray}\label{publicdebt}
D_k=&&(1+r)^k\left[D_0+(g_0-\alpha p_a)\frac{(1+r)^k-1}{r(1+r)^k}\right.\nonumber\\ &&\left.-\gamma a \sum_{i=1}^{k}\frac{b_i^2}{(1+r)^i}-\frac{\beta b_m}{(1+r)^m}\right]
\end{eqnarray}
In the above two equations it is understood that the last term is present only if $k\geq m$. We notice that the first equality in eq. (\ref{Coupon}) is similar to the formula for the price of a coupon bond \cite{Baaquie}.

Therefore, following this model, we can argue that the Government choice of the values of $\alpha$, $\beta$, $\gamma$ and $m$ affects both the average consumer's budget according to Eq. (\ref{consumer}), and the public debt evolution, as specified in (\ref{publicdebt}).

\section{Maximum State expense in paying back public debt}

\begin{figure}[htb]
\includegraphics[scale=0.8]{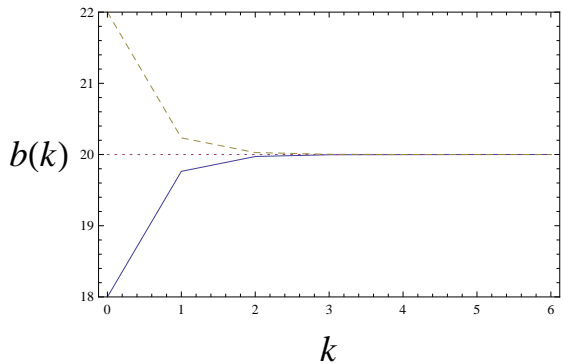}
\caption{Consumer's budget dynamics described by Eq.(\ref{consumer2}) for $\alpha=0.25$, $a=0.15$ and $p_a=100$, for the initial conditions $b_0=18$ (continuous line), $b_0=20$ (dotted line) and $b_0=22$ (dashed line).}
\label{Fig1}
\end{figure}

In this section we study the dynamics governed by Eqs. (\ref{consumer}) and (\ref{publicdebt}) under some specific simplifying assumptions. First of all, we exclude the possibility of direct taxation of the amount $b_k$, so that $\beta=0$. Also, we take $\alpha=\gamma$, since in many countries these rates are in fact quite close. In this way, Eqs. (\ref{consumer}) and (\ref{publicdebt}) get rewritten respectively as
\begin{equation}\label{consumer2}
(1+\alpha)ab_k^2 +b_k-b_{k-1}=(1-\alpha)p_a
\end{equation}
and
\begin{eqnarray}\label{publicdebt2}
D_k=&&(1+r)^kD_0+(g_0-\alpha p_a)\frac{(1+r)^k-1}{r}\\&&\nonumber-\alpha a (1+r)^k\sum_{n=1}^{k}\frac{b_n^2}{(1+r)^2}.
\end{eqnarray}
In Fig.\ref{Fig1} we show the dynamics of the consumer budget described by Eq.(\ref{consumer2}), where we set $a=0.15$ and $p_a=100$, and we assume that $\alpha=0.25$, as is reasonable for an average consumer in a european country. We notice that the fixed point $b_{\lambda}=20$ is reached after few steps.

Let us find the fixed point of Eq.(\ref{consumer2}). The condition $b_{\lambda+1}=b_{\lambda}$ gives
\begin{equation}
b_{\lambda}=\sqrt{\frac{1-\alpha}{1+\alpha}\frac{p_a}{a}}
\end{equation}
Assuming that the average consumer does indeed start with an initial budget close to $b_{\lambda}$, we may rewrite Eq. (\ref{publicdebt2}) as follows
\begin{eqnarray}\label{publicdebt3}
D_k&=&(1+r)^kD_0+(g_0-\alpha p_a)\frac{(1+r)^k-1}{r}\nonumber\\&&\nonumber-\alpha a (1+r)^kb_{\lambda}^2\sum_{n=1}^{k}\frac{1}{(1+r)^n}\\&=& (1+r)^kD_0 + \left(g_0-\frac{2\alpha p_a}{1+\alpha}\right)\frac{(1+r)^k-1}{r}
\end{eqnarray}
The public debt starts decreasing if there is some value of $k$ such that $D_{k+1}<D_k$. This implies the $k-$independent relation
\begin{equation}\label{condition}
p_a>\frac{1+\alpha}{2\alpha}(rD_0+g_0).
\end{equation}
If this condition is met, then, a $k-$independent decrease of the public debt is possible, therefore reducing the debt to GDP ratio, as required by SGP. For example, in a country where $\alpha=1/4$, $r=1/20$, Eq. (\ref{condition}) reads
\begin{equation}\label{condition2}
p_a>\frac{5}{2}\left(\frac{D_0}{20}+g_0\right).
\end{equation}
If $D_0/20\ll g_0$, we have $p_a>2.5 g_0$, which indicates that the value of $g_0$ cannot be higher of $(2/5) p_a$, in order to have decreasing values of $D_k$ for $k>1$.

This result tells that decreasing the salaries and wages of consumers may not be the right choice for a government to reduce the public debt.

\section{The r\^{o}le of the public expenditure}

In this section we consider our model under some more general assumptions. In particular we relax the assumption that the public expenditure $g_k$ is constant. Moreover we consider a more general income-expenditure relation for the consumer, i.e.
\begin{eqnarray}
c_k=a_n b_k^n, \qquad n\geq 2,
\end{eqnarray}
where we excluded a linear behavior in order to model the existence of a threshold. In this case the stable fixed point is given by
\begin{equation}
b_{\lambda}=\sqrt[n]{\frac{1-\alpha}{1+\alpha}\frac{p_a}{a}}.
\end{equation}
Assuming as above that the consumer is in proximity of this fixed point we get, in place of Eq. (\ref{publicdebt3}):
\begin{eqnarray}
D_k&=&(1+r)^{k}D_0-\frac{2\alpha}{(1+\alpha)r}p_a\left[(1+r)^{k}-1\right]\nonumber \\&& + (1+r)^{k}\sum_{i=1}^{k}\frac{g_i}{(1+r)^i}.
\end{eqnarray}
Notice that the assumption of being at the fixed point has again erased the dependence on the coefficient $a_n$. The condition $D_{k+1}<D_k$ in this case gives
\begin{eqnarray}\label{condition3}
\frac{2\alpha p_a}{1+\alpha}> g_1 + r D_0 +\sum_{j=1}^{k-1}\frac{g_{j+1}-g_j}{(1+r)^j},
\end{eqnarray}
where $g_1$ is the expenditure at year 1. Let us now assume that the expenditure in the subsequent years has the form
\begin{eqnarray}
g_j=(j-1)\Delta G  + g_1
\end{eqnarray}
where $g_1>0$. In this way $g_{j+1}-g_j=\Delta G$, so that the sign of $\Delta G$ determines wether the public expenditure grows or decreases over the years. The condition (\ref{condition3}) then specializes to
\begin{eqnarray}\label{condition3}
\frac{2\alpha p_a}{1+\alpha}> g_1 + r D_0 +\Delta G \frac{(1+r)^{k-1}-1}{r(1+r)^{k-1}},
\end{eqnarray}
where $k\geq 1$. For $|\Delta G| \ll g_1$ the last term in the rhs is negligible, therefore our conclusions of the previous section are still valid.

\section{Discussion and conclusions}

In this paper, building on a simple quantitative description of the behavior of an average consumer over a brief period \cite{noi}, and on Barro's theory of the public debt \cite{barro}, we propose a model for the description of the influence of the former on the dynamics of the public debt, with the aim of establishing the condition for its decrease. Our result are clearly limited in scope by the use of the concept of average consumer and by the fact that it is valid only over a short period. We obtained analytical results under the simplifying assumptions of equal average taxation rates for the yearly income and expenditure of the consumer, respectively. Furthermore the possibility of direct taxation of consumer's average yearly bank deposit has been excluded. Besides extending the model to overcome its shortcomings, it would be interesting to further relax the above simplifying assumptions and see when and how the condition we found gets enhanced.

An interesting issue to investigate is how the above results would get modified by generalizing the leaking bucket model for the average consumer behavior \cite{noi}, in such a way to include the effect of different goods on the level of consumer demand.

Another interesting generalization of the present model could be the inclusion of private enterprises as a different agent in the economic landscape.

The proposed model could finally be rephrased in terms of the debt to GDP ratio in order to study the impact of fiscal consolidations on public debt dynamics and how it is related to fiscal multipliers \cite{consolidation1}.

\section*{Author contribution statement}

All the authors contributed equally to the paper.

\end{document}